\documentclass{article}
\usepackage{epsfig}
\usepackage{amssymb}
\usepackage{graphicx}

\begin{document}
\begin{titlepage}
\title{Spin-Exchange in Hydrogen-Antihydrogen Collisions }
\author{A.Yu. Voronin$^{1}$ and P. Froelich$^{2}$ }
\maketitle
\centering{
$^{1}$P.N. Lebedev Physical Institute, \\
53 Leninsky prospect, 117924 Moscow, Russia \\
$^{2}$Department of Quantum Chemistry, Uppsala University, \\
Box 518, SE-75120 Uppsala, Sweden.}
\vskip 30mm

\begin{abstract}
We consider the spin-exchange in ultracold collisions
of  hydrogen ($H$) and antihydrogen ($\bar{H}$) atoms.
The cross sections for transitions between various
hyperfine states   are  calculated.
We show that the process of  spin exchange in $\bar{H}-H$ collisions
is  basically driven by the strong force between  proton and antiproton, that spin-exchange cross-sections are proportional to the difference of singlet and triplet strong force S-wave scattering lengths, while the strong force effects are enhanced about $10^9$ times due to long range atomic forces. We show that any difference in hyperfine structure of $H$ and $\bar{H}$ ( CPT violation ) manifests itself in radical change of spin-exchange cross-sections energy behavior.
\end{abstract}
\end{titlepage}

\section{Introduction}
The successful production of low energy antihydrogen atoms \cite{PhysRevLett.89.233401,amor02}
and  current experiments on their trapping \cite{gabr08,andr08}
have stimulated interest in the interaction of antiatoms with ordinary matter.
The simplest example of matter-antimatter interaction
is given by collisions between  antihydrogen and  hydrogen atoms
\cite{froe02s, froe00,jons01,voro01,armo02,sinh03,froe04a,armo05}. Collisions between  antihydrogen and
helium atoms have also been studied  \cite{annihi2}.  These investigations  revealed that cold collisions between atoms and antiatoms  do not necesserily result  in  annihilation, but may lead to elastic scattering
or  rearrangement reactions.
Interestingly, it was shown  that the  cross sections  for processes that do {\it not}  result  in annihilation (e.g.  elastic scattering) are sensitive to the strong  interaction between particles and antiparticles
\cite{annihi1,annihi2,jons04,armo05,voro08,berg08}. This presents  an  interesting perspective for  investigating   the details
of  strong interactions between matter and antimatter in low energy collisions of atoms and antiatoms.



In the present work we investigate  spin exchange  in hydrogen-antihydrogen collisions.
The oncoming  atoms are assumed to be asymptotically in the  well defined hyperfine states
which can be changed as the result of collisions. This can happen with or without  changing
the  internal energy of the individual atoms so that some of these collisions
appear elastic, apart from spin-flip.
We find that in the case of $H-\bar{H}$ scattering the main source of such spin-exchange transitions is the
{\it strong force}. This is in contrast to spin-exchange in atom-atom collisions (see  \cite{StoofHHSpin,ZygHHSpin} and the references therein),
where the strong-force effects are usually negligible. At the same time the leptonic contribution to spin-exchange in $H-\bar{H}$ collisions  is found negligibly small,
in spite of the fact that the magnetic moments of the leptons (electron ($e$) and   positron ($\bar{e}$)) are on the order of $10^3$ times larger than those for the hadrons (proton ($p$) and antiproton $\bar{p}$)). We will show that  long range van der Waals  interactions
between atoms and antiatoms plays  role of magnifying glass that  exposes the strong force effects on the molecular scale. Another unique feature of the spin-exchange cross sections is their sensitivity to the equality of the hyperfine energy levels in $H$ and $\bar{H}$, ensured by the CPT invariance. Violation of  that invariance results in a radical  change of certain  cross sections for spin-exchange.

The paper is organized as follows. In section \ref{F} we present the formalism  used in calculating   spin-exchange cross sections. Section \ref{R} discusses the main results. The physical sense of these  results is analysed  in terms of a transparent  zero-range potential model of strong forces. Section \ref{C} contains the conclusions and the  Appendix gives  the details of certain derivations.


\section{Formalism}\label{F}
We assume the following four-body Hamiltonian for the $H\bar{H}$ system:

\begin{equation}\label{HSpin}
\widehat{H}=\widehat{H}_{nr}+\sum_\alpha \widehat{W}_\alpha(\textbf{s}_\alpha,\textbf{r}_\alpha).
\end{equation}
Here $\widehat{H}_{nr}$ is the nonrelativistic spin-independent Hamiltonian,  which includes the kinetic
and  Coulomb  interaction energy  for all particles. The subscript $\alpha$ labels the pairs of  interacting particles: $(p\bar{p})$, $(pe)$, $(p\bar{e})$, $(\bar{p}e)$, $(\bar{p}\bar{e})$ or $(e\bar{e})$.
$\widehat{W}_\alpha(\textbf{s}_\alpha,\textbf{r}_\alpha)$ is the spin-dependent pair interaction,
arising from  low order relativistic corrections to the nonrelativisic Hamiltonian;
$\widehat{W}_\alpha(\textbf{s}_\alpha,\textbf{r}_\alpha)$
acts on the spin $\textbf{s}_\alpha$ and  the relative   coordinates $\textbf{r}_\alpha$ of the
particle pair $\alpha$.

The forthcoming   analysis can be greatly  simplified by  taking into account the characteristic spatial and energy scales of the interactions involved in  $H-\bar{H}$ scattering. 
As  pointed out   in \cite{Kolos75,
Shlyap1,vc98,froe02s,voro04,froe00,jons01,zyge04,Arm04}, the rearrangement transitions to Protonium ($Pn$)  and Positronium ($Ps$)
occur  mainly at internuclear
distances smaller  than the rearrangement radius $R_r \simeq 1$ a.u.. For internuclear distances
above $R_r$ the adiabatic approximation, leading to the separation of leptonic and hadronic motion, is well justified. This allows the description of ultralow energy $H-\bar{H}$ scattering within  a   one-channel model based on a nonrelativistic Hamiltonian \cite{voro01,voro04,zyge04}:
\begin{equation}\label{Hmodel}
\widehat{H}_{nr}\simeq \widehat{T}_{kin}+\widehat{V}_{opt}+V_{ad}(R).
\end{equation}
The above  Hamiltonian describes the interatomic motion in a fixed adiabatic leptonic state, where  $\widehat{T}_{kin}$ is the  operator of kinetic energy of internuclear motion,
$V_{ad}(R)$ is the  adiabatic potential in a given leptonic state, and  $\widehat{V}_{opt}$ is the
effective complex optical potential that  accounts for the inelastic rearrangement
and vanishes for interatomic distances $R>R_r$.

In the following we will discuss the ultralow energy collisions of $H$ and $\bar{H}$ in the ground leptonic state. We will use the adiabatic interaction potential $V_{ad}(R)$ obtained in \cite{Stras04}.
It is worth  noting that since the Pauli principle is not forbidding
an electron and a positron to occupy identical spin-space states, the adiabatic potential $V_{ad}(R)$ is independent of the leptonic spin state. This is in contrast to  the case of $HH$, where due to the Pauli principle the adiabatic potential  is  different in singlet and triplet electronic states. This difference is important for our considerations. In fact, the spin-exchange transitions in $H-\bar{H}$ collisions can occur only when the ({\it relativistic}) spin-dependent terms $\widehat{W}_\alpha(\textbf{s}_\alpha,\textbf{r}_\alpha)$ are taken into account. At  the same time in the $HH$ case the main contribution to such transition rates comes from the above   mentioned effective spin dependence of the {\it nonrelativistic} adiabatic potential, while the  explicitly spin-dependent interaction appears as a perturbation. We will show in the following that  the  spin-exchange reactions in $H-\bar{H}$ collisions are  particularly sensitive to the spin dependent strong force between proton and antiproton.

The general form of the spin-dependent interaction is:
\begin{equation}\label{WSpinDep}
\widehat{W}_{\alpha}=W_{\alpha}^0(r_{\alpha})+W_{\alpha}^1(r_{\alpha})(\textbf{L}_{\alpha}\textbf{S}_{\alpha})
+W_{\alpha}^{2a}(r_{\alpha})(\textbf{r}_{\alpha}\textbf{S}_{\alpha})^2+W_{\alpha}^{2b}(r_{\alpha})\textbf{S}^2_{\alpha}.
\end{equation}
In the above formula $\textbf{r}_{\alpha}$, $\textbf{L}_{\alpha}$, $\textbf{S}_{\alpha}$ are relative coordinate, angular momentum and spin operators for the particle pair $\alpha$.

The particular form of the lepton-lepton and lepton-hadron pair interactions can
be obtained  within the QED \cite{Landau4}. In the case of $(pe)$ and $(\bar{p}\bar{e})$ they
are responsible for the hyperfine splitting of the ground state of hydrogen and antihydrogen.
In the present work we will approximate the non-relativistic pair interactions for $(pe)$ and $(\bar{p}\bar{e})$ by the following effective terms \cite{StoofHHSpin}:
\begin{equation}\label{HFint}
\widehat{W}_{(ep),(\bar{e}\bar{p})}=a_{HF}(\textbf{s}_{e,\bar{e}}\textbf{i}_{p,\bar{p}})=\frac{a_{HF}}{2}(\textbf{F}^2-3/2).
\end{equation}
Here $a_{HF}=2.157$ $10^{-7}$ a.u. is the hyperfine constant, $\textbf{s}_{e,\bar{e}}$ is the electron (positron) spin, $\textbf{i}_{p,\bar{p}}$ is the proton (antiproton) spin and $\textbf{F}=\textbf{s}+\textbf{i}$ is the  total spin of the  $(pe)$ or $(\bar{p}\bar{e})$ pair. The above interaction   correctly reproduces    the ground state hyperfine splitting of separated $H$ and $\bar{H}$ atoms.

The influence of  the $(\bar{p}e)$ or $(p\bar{e})$ spin-dependent terms on the spin-exchange reaction rate is expected to be small due to the Coulomb repulsion  that prevents the involved  particles from  approaching  each other. We will neglect such terms in our study.

In the case of $(e\bar{e})$ the spin-dependent  interaction is responsible for the fine structure and annihilation in Positronium \cite{Landau4}. Due to the CP invariance this interaction conserves spin in  Positronium. In our treatment we will use the following simplified model of the $(e\bar{e})$ spin-dependent term:
\begin{equation}\label{Postr}
\widehat{W}_{(e\bar{e})} =\alpha^0 \delta(\textbf{r}_{Ps})\widehat{P}_0+\alpha^1\delta(\textbf{r}_{Ps})\widehat{P}_1.
\end{equation}
Here  $\alpha^0$ and $\alpha^1$ are the singlet and triplet interaction constants:
\begin{eqnarray}
\alpha^0&=&-4.3 \cdot 10^{-4}-i4.9 \cdot 10^{-6} \mbox{ a.u.}\label{alph0},\\
\alpha^1&=&3.3 \cdot 10^{-4}-i4.3 \cdot 10^{-9} \mbox{ a.u.}\label{alph1},
\end{eqnarray}
$\widehat{P}_0=|S_{Ps}=0,0\rangle \langle S_{Ps}=0,0|$ and $\widehat{P}_1=\sum_{M}|S_{Ps}=1,M\rangle \langle S_{Ps}=1,M|$ are  the projection operators on the subspace of singlet or  triplet  states of Positronium.
The suggested form of interaction conserves spin and correctly reproduces the energy shift and annihilation lifetime of  para- and ortho-Positronium. The spin-orbit coupling terms are neglected; we will show below  that such terms are  not important
for the  spin-exchange transition rates in the $H-\bar{H}$ system.

 The leading contribution to the spin-dependent component of the $(p\bar{p})$ interaction arises from  the strong force. Strong interaction is CP invariant and conserves the spin $S_{Pn}$ of Protonium. Furthermore,
this   interaction is attractive and localized to  distances within $R<R_s$ with $R_s\sim 1$ fm ($1.88$ $10^{-5}$ a.u.). It also includes  an absorptive component that is  responsible for  coupling to the  $(p\bar{p})$ annihilation channels.

There are several models of low energy $p\bar{p}$ interaction( see \cite{Carb} and references therein)  that  differ in the detailed behavior of spin-dependent potentials. These models,  however,  give close values for  the $p\bar{p}$ strong force  singlet and triplet scattering lengths $a_{sc}^{0,1}$. Since the range of strong forces is  much shorter compared to  that   of atomic interaction, it is possible to take into account the strong force effect through an  appropriate modification of the boundary condition for the scattering wave function.
Such a procedure is based on  a knowledge of the strong force scattering lengths
only \cite{froe04a,voro08}. Indeed, at distances such that $R_s\ll R\ll R_r$ the interaction is dominated by the Coulomb $p\bar{p}$ potential. The wave function of internuclear motion  at such distances behaves as a superposition of regular $F_0(pR)$ and irregular $G_0(pR)$ Coulomb solutions, with momentum $p\rightarrow 0$ \cite{voro08}. The coefficient that mixes these two solutions,
the $K$-matrix element $k$, is determined by the strong force scattering length
according to \cite{froe04, voro08} :
  \begin{equation}\label{deltStCoul}
  k=\tan(\delta_{sc}^{0,1})=-2\pi M a_{sc}^{0,1},
  \end{equation}
where $M$ stands for the reduced mass of  $p\bar{p}$.
We will use the following values of the $S$-state strong force scattering length \cite{KW,Carb}:
\begin{eqnarray}
a_{sc}^0&=&(1.07-i1.45) \cdot 10^{-5} \mbox{ a.u.}\label{asc0},\\
a_{sc}^1&=&(1.68-i1.06) \cdot 10^{-5} \mbox{ a.u.}\label{asc1}
\end{eqnarray}
The corresponding nuclear $k$-matrix elements become
\begin{eqnarray}
\tan(\delta_{sc}^0)&=&-0.06+i0.08 \label{dscS0},\\
\tan(\delta_{sc}^1)&=&-0.1+i0.06\label{dscS1}.
\end{eqnarray}
In terms of  state vectors $|f(R),S_{Pn}\rangle$, $R_s<R\ll R_r$,  this result can be formulated as follows:
 \begin{eqnarray}
 |f(R),S_{Pn}\rangle \sim  |F_0(pR),S_{Pn}\rangle+\widehat{K} |G_0(pR),S_{Pn}\rangle\\
 \widehat{K}= -2\pi M \left( a_{sc}^0\widehat{\pi}_0+a_{sc}^1\widehat{\pi}_1\right)\label{Kmatr}.
 \end{eqnarray}
 Here the   $\widehat{\pi}_{0,1}$ are   projection operators on    the subspaces of singlet and  triplet  Protonium states. In this way  the effect of strong forces could be taken into account by introducing
the $K$-matrix (\ref{Kmatr}). The latter    determines the boundary condition that can be imposed on
the wave function in the range $R_s<R\ll R_r$.
The effect of the strong force induced  boundary condition (\ref{Kmatr}) is equivalent to the
action of the zero range potential \cite{Demkov}:
\begin{equation}
\label{ZRPstr}
\widehat{V}_s=\frac{2\pi}{M}\left( a_{sc}^0\widehat{\pi}_0+a_{sc}^1\widehat{\pi}_1\right) \delta(\textbf{R})(\frac{\partial}{\partial R}R).
\end{equation}

To proceed with the calculation of spin-exchange rates within the effective adiabatic model one needs to find the contribution of all   spin-dependent terms (\ref{WSpinDep}) to the effective internuclear interaction. We will accommodate these terms  to the first order of perturbation theory. To do so we average the  spin-dependent interactions that  depend on leptonic coordinates over the unperturbed  leptonic wave function $\Psi_{e,\bar{e}}(R,\textbf{r}_{e},\textbf{r}_{\bar{e}})$ (calculated without
the account of spin-dependent interactions). Taking into account our assumptions about the spin-dependent terms and their radial dependence,
such  averaging  will involve only  the  $(e\bar{e})$ interaction:
\begin{eqnarray}
\widehat{V}_{e\bar{e}}(R)=\int|\Psi_{e,\bar{e}}(R,\textbf{r}_{e},\textbf{r}_{\bar{e}})|^2\widehat{W}_{e\bar{e}}d\textbf{r}_{e}d\textbf{r}_{\bar{e}}\\
\widehat{V}_{e\bar{e}}(R)=A(R)\left(\alpha^0 \widehat{P}_0+\alpha^1\widehat{P}_1\right ), \label{VPs}
\end{eqnarray}
where  $A(R)=|\Psi_{e,\bar{e}}(R,\textbf{r}_{e}=\textbf{r}_{\bar{e}})|^2$ is the leptonic coalescence probability. We  use  the  values of the coalescence probability calculated in
ref. \cite{froe04a}. The corresponding spin-dependent leptonic potentials for the singlet and triplet  states
are shown in Fig. 1.
\begin{figure}
  \centering
 \includegraphics[width=115mm]{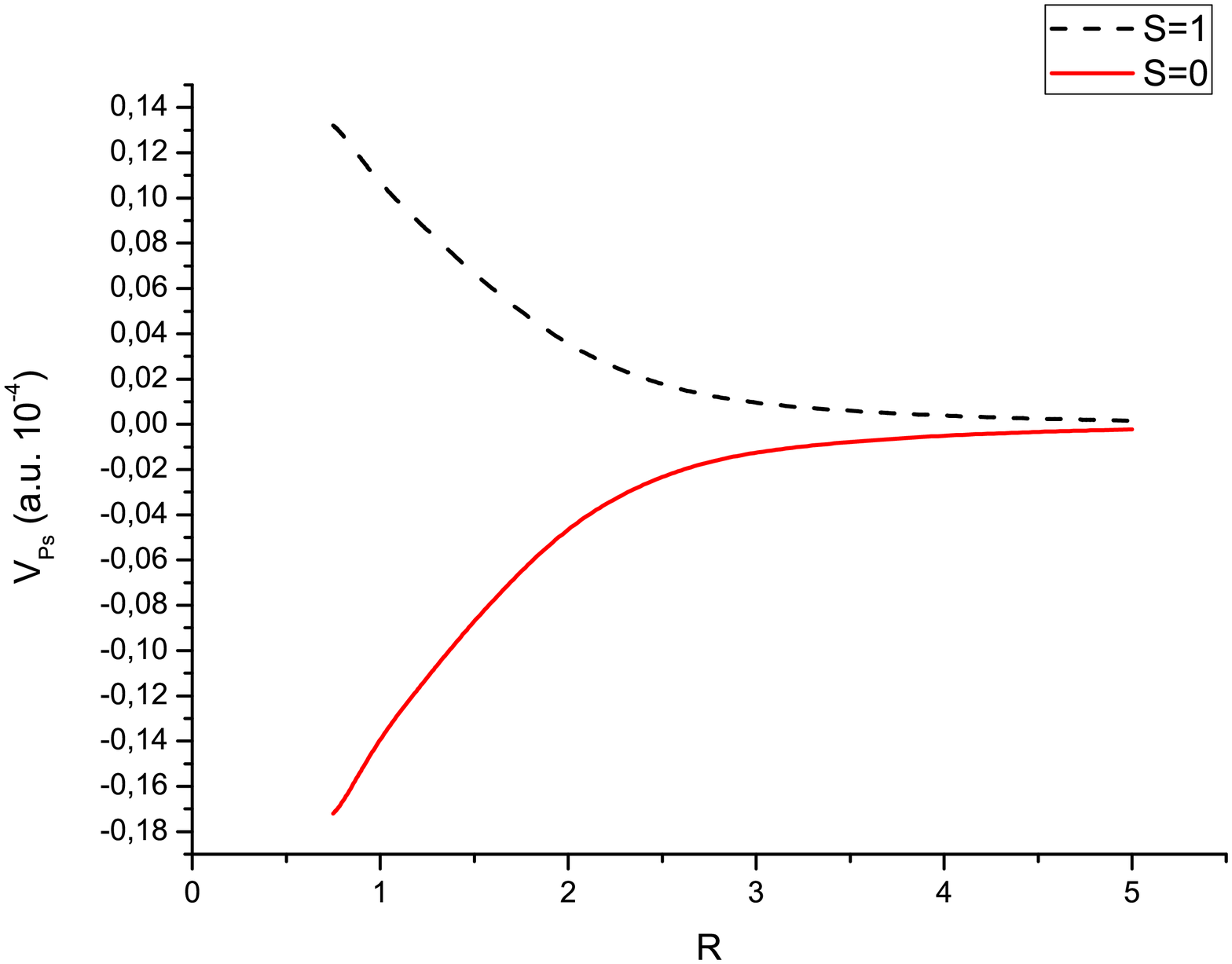}
\caption{Real part of the spin-dependent leptonic potentials for the singlet (dashed line) and
triplet (solid line) states. }
\label{F1}
\end{figure}
We note that the  spin-dependent interactions (\ref{WSpinDep}) also modify the final states of
$Pn$ and $Ps$ in the rearrangement channels, so that the latter may  contribute to the effective optical potential $\widehat{V}_{opt}$. Such a contribution only arises at the level of second order perturbation theory and  will  not be taken into account here.

We may now   formulate the scattering problem for spin-exchange transitions. The spin state of the colliding $H-\bar{H}$ system   can be described in different representations, in terms of $16$ spin basis vectors.  Since  the spin-dependent terms for $(p\bar{p})$ and $(e\bar{e})$ interactions  conserve spin of the corresponding pairs,  they will be  diagonal in the basis set with given $(p\bar{p})$ and $(e\bar{e})$ spins. Such basis set will be denoted  $|S_{(p,\bar{p})},M_{(p,\bar{p})},S_{(e,\bar{e})},M_{(e,\bar{e})}\rangle$. We will  hereafter refer to
this basis as the $S$-representation.

The asymptotic Hamiltonian,   which describes the separated $H$ and $\bar{H}$ atoms and  includes the hyperfine terms  $\widehat{W}_{(pe)}$ and $\widehat{W}_{(\bar{p}\bar{e})}$, is diagonal in the basis with given $(pe)$ and $(\bar{p}\bar{e})$ spins  $F_H$ and $F_{\bar{H}}$, respectively. Such basis set will be denoted  $|F_H,M_H,F_{\bar{H}},M_{\bar{H}}\rangle$ and will be  called the $F$-representation. The scattering state vector $|\varphi \rangle$ is a superposition of basis spin states:
\begin{equation}
|\varphi\rangle=\sum_{F_{H},M_{H},F_{\bar{H}},M_{\bar{H}}}f(R)_{F_{H},M_{H},F_{\bar{H}},M_{\bar{H}}}|F_H,M_H,F_{\bar{H}},M_{\bar{H}}\rangle
\end{equation}
The $R$-dependent expansion coefficients play  the role of  hadronic wave functions for the
 $H\bar{H}$ system in a given spin state of four particles. The asymptotic form of
functions $f(R)$ at large $R$ gives in a standard way the $S$-matrix in the given spin-state representation.
The desired  wave functions are found from the solution to  the coupled equation system:
\begin{equation}\label{eqSys}
 \sum_{\mu}\langle\mu'|\widehat{H}-E|\mu\rangle F(R)_{\mu,\eta}=0,
\hskip 10mm \eta = 1,2, ..., 16
\end{equation}
Here $F(R)_{\mu',\mu}$ is a $16\times16$ solution matrix whose indices  label the outgoing
($\mu'$) and  incoming ($\mu$) channels, respectively; and $\eta$ labels  the sets of spin quantum numbers $F_H,M_H,F_{\bar{H}},M_{\bar{H}}$.

While solving the coupled channel equations   we utilize the fact  that for interatomic distances $R\sim 1$ a.u. and below the characteristic values of the local adiabatic potential are much greater than the hyperfine splitting.    Thus even in the
zero-energy limit the  hyperfine terms may be neglected in the solution
(\ref{eqSys}) at  short and intermediate internuclear distances.
At such distances the equation system can be decoupled in the $|S_{(p,\bar{p})},M_{(p,\bar{p})},S_{(e,\bar{e})},M_{(e,\bar{e})}\rangle$ basis:
\begin{equation}\label{DiagEQS}
\left(-\frac{1}{2M}\frac{d^2}{dR^2}+\widehat{V}_{opt}+V_{ad}(R)+A(R)\alpha^{S_{(e\bar{e})}}-E\right)F(R)_{\mu,\mu} = 0,
\end{equation}
where  $\mu$ denotes  a set of quantum numbers ${S_{(p,\bar{p})},M_{(p,\bar{p})},S_{(e,\bar{e})},M_{(e,\bar{e})}}$.
Such an equation system should be supplied with boundary conditions, imposed  at
arbitrary $R_0$ within  the range $R_s\ll R_0\ll R_r$, and specified  by the strong force $K$-matrix (\ref{Kmatr}):
\begin{eqnarray}\label{DiagBoundCond}
F(R_0)_{\mu,\mu'} = \delta_{\mu,\mu'}\left(F_0(p R_0)-2\pi a_{sc}^{S_{(p\bar{p})}}G_0(p R_0)\right)\\
\label{DiagBoundCond2}
F'(R_0)_{\mu,\mu'} = \delta_{\mu,\mu'}\left(F'_0(p R_0)-2\pi a_{sc}^{S_{(p\bar{p})}}G'_0(p R_0)\right).
\end{eqnarray}

Instead of solving the equation system (\ref{DiagEQS}) with the boundary conditions
(\ref{DiagBoundCond}-\ref{DiagBoundCond2}) we  use the  more practical method developed in \cite{voro08}.
It amounts to introducing  specially fitted  boundary conditions  at the \emph{rearrangement radius} $R=R_r$:
\begin{equation}\label{deltaBound}
\left(F'(R_r)F^{-1}(R_r)\right)_{\mu,\mu'}
=\delta_{\mu,\mu'} p(R_r)\cot(\delta+\delta_{sc}^{S_{(p\bar{p})}})
\end{equation}
where
$p(R_r)=\sqrt{2M V_{ad}(R_r)}$ is a classical local momentum given
at the internuclear separation $R_r$.
The phase shift $\delta$ takes into  account the  effect of the optical and adiabatic potentials accumulated at distances below $R_r$. The account of strong forces is achieved by adding to the phase shift $\delta$  the spin dependent strong force phase shift $\delta_{sc}^{S_{(p\bar{p})}}$. The additivity of phases is explained by the fact that contribution of the optical potential  to the phase shift is vanishing  in the region $R<R_s$, characteristic for strong forces. The phase shift $\delta$ is calculated using the model nonlocal optical potential
\cite{voro08} and is given by:
\begin{equation}\label{deltaCV}
\delta=0.74+i0.32.
\end{equation}
The advantage of the above method is that once the phase shift $\delta$ is calculated, the effect of rearrangement transitions is incorporated through the boundary condition (\ref{deltaBound}) obviating cumbersome calculations with nonlocal potentials. The  phase shift approach gives a useful  universal tool for the treatment and analysis of rearrangement transitions.

Once the (decoupled) equation system is solved in the $S$-representation, the solutions (valid in the
limited range below $R \sim 1$ a.u.) should be transformed back to the   asymptotically correct $F$-representation.  This is done by means of a unitary transformation \cite{ZygHHSpin}:

\begin{eqnarray}\label{transform}
\widetilde{F}(R)= U^+F(R)U\\
U=\langle S_{(p,\bar{p})},M_{(p,\bar{p})},S_{(e,\bar{e})},M_{(e,\bar{e})}|F_H,M_H,F_{\bar{H}},M_{\bar{H}}\rangle \label{U}
\end{eqnarray}

Thus obtained solution $\widetilde{F}(R)$ should be matched at some point $R_0\sim 1$ a.u. with a solution $\Phi$, in   which now the hyperfine energy splitting is taken into account:
\begin{eqnarray}\label{eq)}
\sum_{\mu }\left[\left(-\frac{1}{2M}\frac{d^2}{dR^2}+V_{ad}(R)\right)\delta_{\mu',\mu}+[U^+ \widehat{V}_{(e\bar{e})}U]_{\mu',\mu}+Q_{\mu',\mu}\right]\Phi_{\mu,\eta}(R)=E\Phi_{\mu',\eta}\\
\left[\widetilde{F}^{\prime}(R_0)\widetilde{F}^{-1}(R_0)\right]_{\mu',\mu}
=\left[\Phi^{\prime}(R_0)\Phi^{-1}(R_0)\right]_{\mu',\mu}. \label{eqBC}
\end{eqnarray}
Here $Q_{\mu',\mu}$ is a diagonal ($16\times16$) hyperfine energy matrix, which gives the threshold energies in channels with different combinations of total spin $F_{H}$ and $F_{\bar{H}}$. Its diagonal values are:

\begin{equation}
Q_{\mu,\mu}=\alpha_{HF}\left\{
\begin{array}{clll}
0 & \mbox{if} & (F_H=0,F_{\bar{H}}=0) \\
1 & \mbox{if} & (F_H=1,F_{\bar{H}}=0) \mbox{ or } (F_H=0,F_{\bar{H}}=1)\\
2 & \mbox{if} & (F_H=1,F_{\bar{H}}=1)
\end{array}
\right.  \label{Qmatr}
\end{equation}

From the conservation of the projection of total spin ($M_{H\bar{H}}=M_{H}+M_{\bar{H}}$)  it follows that the $\Phi$-matrix has a block structure, consisting of a $6\times6$ matrix ( $M=0$) , two $4\times4$ matrixes ($M_{H\bar{H}}=1$ and $M_{H\bar{H}}=-1$) and two $1\times1$ matrixes ($M_{H\bar{H}}=2$ and $M_{H\bar{H}}=-2$).

The such obtained  solution matrix  has the following asymptotic form as $R\rightarrow \infty$:
\begin{equation}
\Phi(R)\rightarrow \exp(-i\widehat{p}R)-\exp(i\widehat{p}R)\widehat{p}^{-1/2}\widehat{S}\widehat{p}^{1/2}.
\end{equation}
Here $\widehat{p}$ is a diagonal matrix whose elements are $\widehat{p}_{\mu,\mu}=\sqrt{2M(E-Q_{\mu,\mu})}$; $\exp(\mp i\widehat{p}R)$ is a diagonal matrix with elements $[\exp(\mp i\widehat{p}R)]_{\mu,\mu}=\exp(\mp i\widehat{p}_{\mu,\mu}R)$;  and  $\widehat{S}$ is the desired $S$-matrix.

\section{Results and discussion}\label{R}
We consider   the  $H-\bar{H}$ collisions in the absence of an external magnetic field  ($B=0$). We  follow the usual notation for the hyperfine states, i.e. the $H$ state $|F_H=0,M_H=0\rangle$ is denoted as $a$, the state $|F_H=1,M_H=-1\rangle$ is denoted $b$, the state $|F_H=1,M_H=0\rangle$ is denoted as $c$ and the state $|F_H=1,M_H=1\rangle$ is denoted $d$. The corresponding  spin-states of $\bar{H}$  are denoted  $\bar{a}$, $\bar{b}$, $\bar{c}$ and $\bar{d}$. We present three types of typical  behavior of spin exchange cross sections as a function of collision energy. The spin-exchange cross sections $\sigma_{d\bar{a}\rightarrow a\bar{d}}$ and $\sigma_{d\bar{a}\rightarrow c\bar{d}}$ are shown on Fig. 2; the cross section $\sigma_{c\bar{d}\rightarrow d\bar{a}}$ is shown on Fig. 3.

\begin{figure}
  \centering
 \includegraphics[width=115mm]{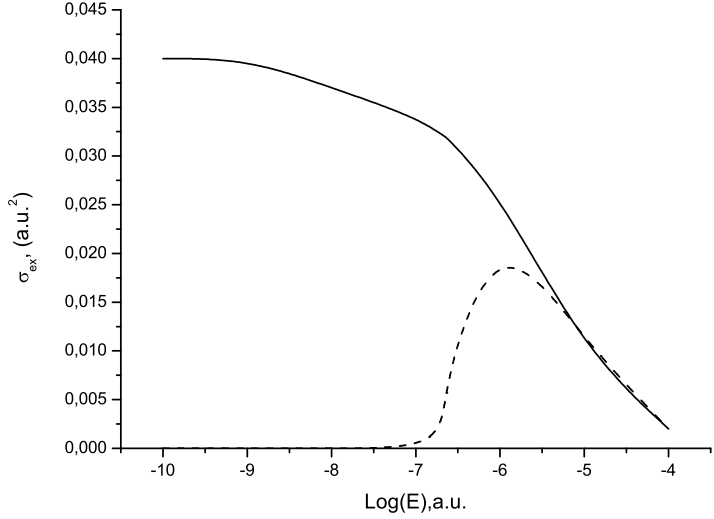}
\caption{The spin-exchange cross sections $\sigma_{d\bar{a}\rightarrow a\bar{d}}$ (solid line) and $\sigma_{d\bar{a}\rightarrow c\bar{d}}$ (dashed line).}
\label{F2}
\end{figure}

\begin{figure}
  \centering
 \includegraphics[width=115mm]{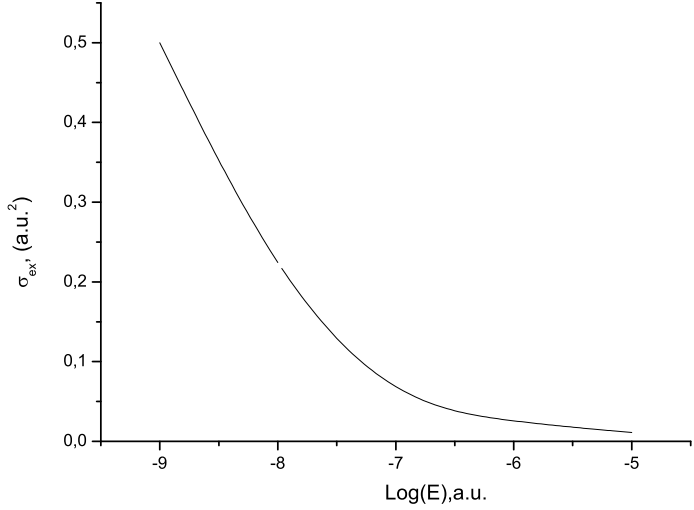}
\caption{The spin-exchange cross section $\sigma_{c\bar{d}\rightarrow d\bar{a}}$.}
\label{F3}
\end{figure}

   According to the CPT invariance,  the energies of the $d\bar{a}$ and $a\bar{d}$ states  are equal and so the reaction $d\bar{a}\rightarrow a\bar{d}$ takes place even in the limit of vanishing collision energy, $E\rightarrow 0$. In that limit the cross section tends to a  constant value,
$\sigma_{d\bar{a}\rightarrow a\bar{d}}(E)\rightarrow 0.04$ a.u.$^2$.

The reaction $d\bar{a}\rightarrow c\bar{d}$ can only occur above the threshold  energy $Q_{d\bar{a}\rightarrow c\bar{d}}=\alpha_{HF}$. The   threshold behavior of the associated cross section is clearly seen on Fig. 2. Again, according to the CPT invariance, the cross sections for  conjugated reactions are  the same, giving  $\sigma_{d\bar{a}\rightarrow c\bar{d}}=\sigma_{a\bar{d}\rightarrow d\bar{c}}$.
 The inverse reaction $c\bar{d}\rightarrow d\bar{a}$ is exothermic (in our case the energies of the final states for  exothermic reactions are smaller than the energies of the initial states
by $\alpha_{HF}$ or $2\alpha_{HF}$) so that  the cross section varies  as  $1/v$, where $v$ is the incident channel velocity. Such  behavior is shown in Fig. 3. Hence, for collisional energies $E<\alpha_{HF}$  the cross sections of exothermic reactions are the largest among  spin-exchange  cross sections.
Let us mention that a  hypothetical violation of CPT invariance  would manifest itself in a  radical change of  the spin-exchange  cross sections. In particular, the energy difference between the hyperfine levels of $H$ and $\bar{H}$ (induced by the CPT violation)  would result in  the appearance of  new reaction  thresholds. Consequently,   the cross sections of the reactions $d\bar{a}\leftrightarrows a\bar{d}$ would follow  the above mentioned ($v$ and $1/v$)  threshold pattern, instead of tending to the established constant value in the limit $E\rightarrow 0$.

 For increasing collision energies, i.e. when  $E\gg \alpha_{HF}$, all  spin-exchange cross sections tend to the same limit. In such a case the hyperfine splitting can be neglected even in the asymptotic states. Therefore the  $S$-matrix is diagonal in $S$-representation. The $S$-matrix in the $F$-representation (needed in our   calculation of   cross sections) is obtained by the transformation specified in eq. (\ref{U}). One can show (using the explicit form of the transformation matrix $U$ \cite{ZygHHSpin}) that  the $S$-matrix elements in the $F$-representation for the reactions $d\bar{a}\rightarrow a\bar{d}$  and $d\bar{a}\rightarrow c\bar{d}$, are connected to  the matrix elements in the $S$-representation  in the following way:
 \begin{eqnarray}\label{Sij1}
 S_{d\bar{a}\rightarrow a\bar{d}}&=&\frac{(\langle 0,1| S|0,1\rangle+\langle 1,0| S|1,0\rangle)}{4}-\frac{\langle 1,1| S|1,1\rangle}{2},\\
 S_{d\bar{a}\rightarrow c\bar{d}}&=&\frac{(\langle 0,1| S|0,1\rangle-\langle 1,0| S|1,0\rangle)}{4}\label{Sij2}.
 \end{eqnarray}
  Here $\langle i,j| S|i,j\rangle$ is the $S$-matrix element in  the $S$-representation, with $S_{p\bar{p}}=i$ and $S_{e\bar{e}}=j$  ($i,j=0,1$). It follows from (\ref{Sij1}, \ref{Sij2}) that the equality of cross sections $\sigma_{d\bar{a}\rightarrow a\bar{d}}=\sigma_{d\bar{a}\rightarrow c\bar{d}}$ requires:
  \begin{equation}\label{S1011}
  \langle 1,0| S|1,0\rangle=\langle 1,1| S|1,1\rangle.
  \end{equation}

The above  equality means that  the elastic scattering in the  state
$|S_{p\bar{p}}=1,S_{e\bar{e}}=0\rangle$ is the same as in the state $|S_{p\bar{p}}=1,S_{e\bar{e}}=1\rangle$. This can hold if the contribution of the ($e\bar{e}$) spin-dependent potential (\ref{VPs}) is negligible. A  calculation made without the inclusion of the potential (\ref{VPs}) fully confirmed this assertion. The scattering lengths $a_{10}$ and $a_{11}$ in the states $|S_{p\bar{p}}=1,S_{e\bar{e}}=0\rangle$ and $|S_{p\bar{p}}=1,S_{e\bar{e}}=1\rangle$ turned out to be:
  \begin{eqnarray}
  a_{10}=5.665-i2.216 \mbox{ a.u.}\\
  a_{11}=5.677-i2.216 \mbox{ a.u.}
  \end{eqnarray}
  while the  scattering length $a_{1}$, calculated  without inclusion of the potential (\ref{VPs}) turned out to be:
  \[a_{1}=5.666-i2.216 \mbox{ a.u.}.\]
  The small  difference between these scattering lengths proves that the role of
spin-dependent $e\bar{e}$ interaction is negligible. To understand this
we note  that the  ratio  $|V_{e\bar{e}}(R)/V_{ad}(R)|$ never exceeds $10^{-3}$ and reaches a maximum in  the range  $3$ a.u.\,$<R<5$ a.u.. At these distances the WKB approximation is still justified. The semiclassical phase accumulated  at these distances by the wave function (without account of the $V_{e\bar{e}}(R))$ is:
  \[\delta_{WKB}=\int_{R_1}^{R_2} \sqrt{2M|(V_{ad}(r)|}dr.  \]
  The change of  that  phase due to the account  of spin-dependent leptonic potential  $V_{e\bar{e}}(R)$ is :
  \[\Delta\delta_{WKB} =\int_{R_1}^{R_2} \frac{\sqrt{M}V_{e\bar{e}}(R)}{\sqrt{2|V_{ad}(R)|}}dr < 10^{-3} \frac{\delta_{WKB}}{2} \]
 which explains the small contribution of the leptonic spin-dependent interaction to the  phase of  the wave function,  and therefore  to the $S$-matrix \cite{voro08}.
Thus we come to the  conclusion that the rate of spin-exchange in
$H-\bar{H}$ collisions is determined \emph{by the $p\bar{p}$ strong interaction}.

The short-range character of strong forces enables the factorization of  nuclear and atomic interactions and simplifies the study of the  energy dependence of the spin-exchange cross sections.  At short distances, where spin transitions take place, the energy difference between the  final states can be totally neglected. Using the explicit form of the transition matrix $U$ (\ref{U}) and the zero-range pseudo-potential (\ref{ZRPstr}) one can get the transition amplitude between  different spin states in the distorted wave approximation:
\begin{equation}\label{Fex}
f_{\alpha \beta}=\frac{a_{sc}^1-a_{sc}^0}{4}\int \Psi_{\beta}^{0*}(\textbf{R})\delta(\textbf{R})(\frac{\partial}{\partial R}R)\Psi_{\alpha}^{0}(\textbf{R})d^3R=\frac{a_{sc}^1-a_{sc}^0}{4}\Psi_{\beta}^{0*}(0)\Psi_{\alpha}^{0}(0).
\end{equation}
Here  $\Psi_{\alpha,\beta}^0(\textbf{R})$ is the  wave function corresponding the initial (final) spin-state $\alpha$ ($\beta$), calculated without taking strong forces into account. (We note that the exact expression for the transition amplitude can be obtained from (\ref{Fex}) upon  replacing  $\Psi_{\alpha,\beta}^0(\textbf{R})$ by  the wave function calculated with the account of strong forces.)  The only contribution to the spin-exchange amplitude arises from the  $S$-wave, since the  higher partial wave functions tend to zero at the origin. The spin exchange cross section is then
given by the following expression:
\begin{equation}\label{CrSec}
\sigma_{\alpha\beta}=\pi\frac{p_{\beta}}{p_{\alpha}}\frac{|a_{sc}^1-a_{sc}^0|^2}{4}|\Psi_{\beta}^0(0)\Psi_{\alpha}^0(0)|^2,
\end{equation}
where $p_{\alpha,\beta}$ is the channel momentum.
It is  convenient to perform a further analysis in terms of the Jost function $f_J$. Following  the definition of the $S$-wave Jost function \cite{Newton} we may for our case write
\begin{equation}\label{Jost}
\Psi_{\alpha}^0(\textbf{R})= \frac{\phi_{\alpha}^0(R)}{R f_J(p_\alpha)}
\end{equation}
where $\phi_{\alpha}^0(R)$ is the regular   $S$-wave  radial solution of the
Schr\"odinger equation (neglecting  strong forces) in spin-state $\alpha$. The behavior of the regular solution at the origin is
\[\phi_{\alpha}^0(R\rightarrow 0)=R \]
so that
\begin{equation}\label{Psi0}
\Psi_{\alpha}^0(0)= \frac{1}{ f_J(p_\alpha)}.
\end{equation}
Equation (\ref{CrSec}) now takes the form:
\begin{equation}\label{CrSecJost}
\sigma_{\alpha\beta}=\pi\frac{p_{\beta}}{p_{\alpha}}\left|\frac{a_{sc}^1-a_{sc}^0}{2f_J(p_{\beta})f_J(p_{\alpha})}\right | ^2.
\end{equation}
Near the threshold the momentum dependence of the Jost-function is known to be \cite{Newton}:
\begin{equation}\label {JostP}
f_J(p_\alpha)=f_J(0) \exp(-i p_\alpha a)
\end{equation}
where $a$ is the atom-antiatom scattering length (without the account of strong forces).
Numerical calculations give the following values for  $a$ and   $1/|f_J(0)|^2$:
\begin{eqnarray}
\label{a}
a&=&5.2-i1.8 \mbox { a.u.}\\ \label{FJ0}
1/|f_J(0)|^2&=&29067.
\end{eqnarray}
We can exploit the smallness of the energy thresholds (determined by the hyperfine constant $\alpha_{HF}$) in different spin states,  and get for the near-threshold behavior of the spin-exchange cross section:
\begin{equation}\label{SigmaEx}
\sigma_{\alpha\beta}=\pi\frac{p_{\beta}}{p_{\alpha}}\exp(2\mathop{\rm Im} a (p_\alpha+p_\beta))\left|\frac{a_{sc}^1-a_{sc}^0}{2f^2_J(0)}\right |^2.
\end{equation}

It follows from the above formula that,  in the case  $p_\alpha=p_\beta$, the cross section in the zero-energy limit tends to the constant value:
\begin{equation}\label{S0}
\sigma_0=\pi\left|\frac{a_{sc}^1-a_{sc}^0}{2f^2_J(0)}\right |^2=0.04 \mbox{ a.u.}
\end{equation}
while in the  presence  of  the energy threshold the behavior of the cross section is determined by the imaginary part $\mathop{\rm Im}a$ of the  atom-antiatom scattering length:
\begin{equation}\label{SigmaEx1}
\sigma_{\alpha\beta}=\frac{p_\beta}{p_\alpha}\exp(2\mathop{\rm Im} a (p_\alpha+p_\beta))\sigma_0.
\end{equation}
Comparison of the above expression, obtained in the distorted wave approximation, with the results of numerical calculations, shows that the formula (\ref{SigmaEx1}) is accurate to within a few percent for  collision energies below $10^{-5}$ a.u.

The behavior of the cross section (\ref{SigmaEx1}) near the threshold has a clear physical meaning. Spin-exchange can only occur  when the nuclei approach distances characteristic for strong forces, i.e. $1$ fm. Due to  inelastic rearrangement transitions into the $Pn+Ps$ channels the flux of $H-\bar{H}$ at such distances is damped by the factor $\exp(2\mathop{\rm Im} a (p_\alpha+p_\beta))$. The damping rate is given by the imaginary part of  scattering length for $H-\bar{H}$ collisions. We note  that the imaginary part  of the scattering length can be extracted  from the  energy dependence of the spin-exchange cross sections \cite{Prot}.

In the case of  exothermic reactions (e.g.  $c\bar{d} \rightarrow d\bar{a}$ or $b\bar{d}\rightarrow a\bar{a}$) there is an energy excess in the final channel: $\Delta E_{\beta}=n \alpha_{HF}$,  $n=1,2$.
It follows from (\ref{SigmaEx1}) that the  cross sections for such reactions behave like $d_n/v$ in the limit $E\rightarrow 0$. Taking into account that as  $ E\rightarrow 0$
$p_{\beta} \rightarrow \sqrt{2M n \alpha_{HF}}$,    we obtain  for $d_n$:
\begin{equation}\label{Dn}
d_n=\frac{\sqrt{2M n \alpha_{HF}}}{M}\exp(2\mathop{\rm Im} a\sqrt{2M n \alpha_{HF}} )\sigma_0,
\end{equation}
which gives:
\begin{eqnarray}
d_1&=&8.0\times 10^{-7} \mbox{ a.u.}^2 \\
d_2&=&1.1\times 10^{-6} \mbox{ a.u.}^2
\end{eqnarray}

The ratio of  the near-threshold spin-exchange  reaction rates  $b\bar{d}\rightarrow a\bar{a}$ and $c\bar{d} \rightarrow d\bar{a}$   is given by:
\begin{equation}\label{Dratio}
d_2/d_1=\sqrt{2}\exp\left(2(\sqrt{2}-1)\mathop{\rm Im} a\sqrt{2M  \alpha_{HF}}\right)=1.375.
\end{equation}
As one can see, this ratio is determined by the product $\mathop{\rm Im} a\sqrt{2M  \alpha_{HF}}$.

The cross sections of the inverse reactions ($d\bar{a} \rightarrow c\bar{d}$ or $a\bar{a}\rightarrow b\bar{d}$) show a characteristic  threshold behavior. In the vicinity of the threshold they
 behave like $pd_n/\sqrt{2M n \alpha_{HF}}$.

It was already mentioned that for  $E\gg \alpha_{HF}$ all the spin-exchange cross sections tend to the same limit. Within  the zero-range potential model of strong forces, this limit  is given by:
\begin{equation}\label{largeELim}
\sigma=\exp(4\mathop{\rm Im} a p) \sigma_0.
\end{equation}

A remarkable feature of the obtained results is the factorization of the
spin-dependent strong force  effect and the effect of long-range atomic interaction (\ref{SigmaEx}). While the first  appears through the difference between singlet and triplet \textit{strong force} scattering lengths, the second  is given by the Jost function for the  interatomic motion of the $H-\bar{H}$ pair.  This latter factor ($|1/f_J(0)|^4$) strongly enhances (approximately $ 10^9$ times!) the effect of strong forces, due to focusing of the flux of $p-\bar{p}$ towards the center by the molecular potential. The large value of the enhancement  factor is determined by the large ratio between  atomic  and nuclear scales.

Under certain conditions, the enhancement factor $|1/f_J(0)|^4$ could become even larger. This additional enhancement is due to the  existence of the near-threshold quasi-molecular  $H\bar{H}$ states \cite{voro08}.
Indeed, such quasi-bound states manifest themselves as zeros of the Jost function $f_J(p)$ in the complex momentum plane.  The presence of these  zeros in the threshold vicinity leads to a  small value for  $f_J(0)$. To study the question in more detail we will use the analytical expression for the Jost function  based on the WKB approximation for  interatomic motion at
distances $1$ a.u. $<R<5$ a.u. (see Appendix):
\begin{equation}\label{JostWKB}
\frac{1}{|f^{WKB}_J(0)|^2}=\frac{2\pi M a_0}{\sin^2(\pi/8+\Omega+\delta)}.
\end{equation}
In the above equation, $a_0$ is the characteristic scale for van der Waals  interaction between
$H$ and $\bar{H}$ \cite{voro08}:
\begin{equation} \label{a0}
a_0=\sqrt[4]{2MC_6}\frac{\Gamma (3/4)}{2\sqrt{2}\Gamma (5/4)}\simeq
4.99\mbox{ a.u.}
\end{equation}
$C_6=6.5$ is the van der Waals  constant for $H\bar{H}$ interaction, $\Omega$ is the semiclassical phase, accumulated in the distance range $R> R_r$
\begin{equation}\label{Omega}
\Omega=\int_{R_r}^\infty\sqrt{2MV_{ad}(R)}dR=19.38
\end{equation}
and $\delta=0.74+i 0.32$ is the \textit{complex} phase shift (\ref{deltaCV}) accumulated at distances less than $R_r$.
The value of expression (\ref{JostWKB}), obtained within the WKB approximation,  turns out to be
 \begin{equation}\label{valueJostWKB}
\frac{1}{|f^{WKB}_J(0)|^2}=24\,830
\end{equation}
and is comparatively close to the exact value (\ref{FJ0}).

 Following the method developed in \cite{voro08}, we will vary the total  phase of the wave function  $\varphi=\pi/8+\delta+\Omega$. The physical sense of such  procedure is the following. It is known that interatomic interactions can be modified by means of external magnetic or electric fields
(see \cite{FieldE, FieldM} and references therein). Without  specifying the physical method of such  modification  we may describe the resulting effect as a variation of the accumulated phase of the wave function. We will be interested in such modification of interatomic interaction that results in the  increase of the enhancement factor (\ref{FJ0}).
This factor (\ref{FJ0}) would become infinite (the Jost function $f_J(0)=0$) if there is a quasi-molecular state exactly at the threshold. In such a case $\varphi=\pi k$, $k=0,1,2...$. As long as there is an imaginary component of the total phase $\varphi$ (which represents absorption of the $H-\bar{H}$ flux into rearrangement channels) such  a condition can not be satisfied exactly; however the enhancement factor will show  maxima corresponding to  values of the total phase satisfying $\mathop{\rm Re}\varphi=\pi k$. As was shown in \cite{voro08} this condition corresponds to the appearance of decaying quasi-bound state in the spectrum of the $H\bar{H}$ system close to the dissociation threshold.

\begin{figure}
  \centering
 \includegraphics[width=115mm]{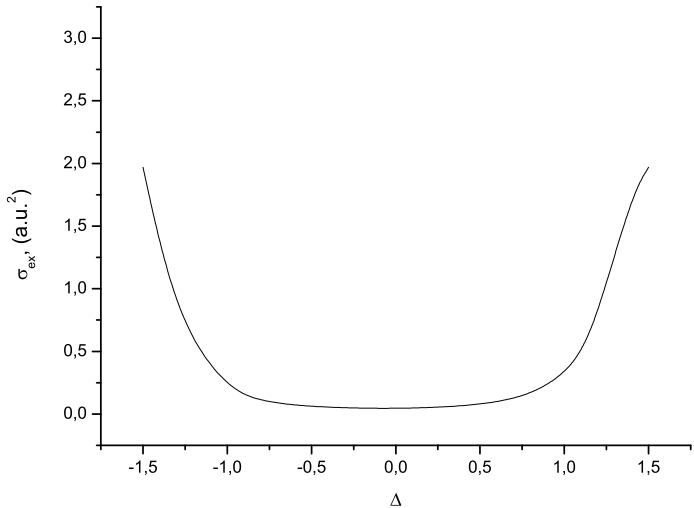}
\caption{The zero-energy limit of the $c\bar{d}\rightarrow d\bar{c}$  cross section as a  function of the phase variation that expresses a modification of the interatomic interaction due to the presence of an external field.}
\label{Enh}
\end{figure}

In Fig. \ref{Enh} we present the zero-energy limit of the  cross section for the reaction  $c\bar{d}\rightarrow d\bar{c}$  as a function of phase variation. For certain values of the  total phase $\varphi$ the spin-exchange cross section increases more than 50 times in comparison with
its "unbiased" value ($\varphi=20.51+i 3.2$) obtained in the absence of any external perturbation. Strictly speaking, in such a resonant case the distorted wave approximation (\ref{Fex}) is not justified and one should substitute the approximate value of $\Psi_{\alpha,\beta}^0(0)$ by the wave function calculated with the account of strong forces. Nevertheless, this result indicates  that the modification of the interatomic $H-\bar{H}$ interaction by external fields could in principle strongly increase the spin-exchange cross sections, exposing the effect of strong forces.
Such additional enhancement might also occur in collisions with other atoms, due to the sensitivity of the cross sections to the reduced mass of the colliding nuclei
\cite{voro08}.
\section{Conclusions}\label{C}
We found that the  spin-exchange transitions in $H-\bar{H}$ collisions are caused mainly by the strong force. This mechanism of spin-exchange is  a unique feature of atom-antiatom interaction, with no  counterpart in atom-atom collisions. The corresponding transition amplitudes are proportional to the difference between  singlet and triplet strong force scattering lengths. Meanwhile, the cross sections acquire  molecular dimensions  due to the large  enhancement of the strong force effect; this enhancement originates from the  presence of long range,  atom-antiatom  interaction. We  have shown  that the  enhancement factor is on the order of  $10^9$.

The energy behavior of the spin-exchange cross sections turns to be sensitive to the difference between hyperfine energy levels  in  $H$ and $\bar{H}$ (which might occur due to CPT violation). If such  violation were to occure the threshold behavior of the spin-exchange cross sections would be radically different from the CPT invariant case. We present a universal description of   the influence of external fields on the  $H-\bar{H}$ scattering,
which suggests that  field-induced modification  of the interatomic interaction can additionally enhance (up to 50 times) the spin-exchange cross sections, exposing the strong force effects.

\section*{Acknowledgments}
We  would like to acknowledge the support from the
Swedish  Research Council, from the  Wenner-Gren
Foundations and  the Royal Swedish Academy of Sciences. One of the authors (A.V.) would like to thank E. Shulgina for very useful discussions and comments.

\section{Appendix}
The present Appendix outlines the derivation of the  analytical expression for
the Jost function $f_J(0)$ for a  zero collision energy of the $H-\bar{H}$ system.
We start with the expression for the $S$-wave  scattering length for $H-\bar{H}$, obtained in \cite{voro08} under the following assumptions. At distances $R>R_r\sim 1$ a.u. the $H-\bar{H}$ interaction is well described by the one-channel, adiabatic potential $V_{ad}(R)$. At distances $R>R_{h}\sim 5$ a.u. this  potential can be accurately reproduced by the van der Waals interaction  $-C_6/R^6$, with $C_6=6.5$. There exists a domain $R_r<R<R_h\ll R_{vdW}$, with $R_{vdW}=\sqrt[4]{2M C_6}\simeq 10.5$ a.u., where the WKB approximation is well justified.

Matching the WKB wave function to the exactly known asymptotic zero energy solution of the Schr\"{o}dinger equation with  the $-C_6/R^6$ potential,  one gets the following expression for the scattering length in  $H-\bar{H}$ collisions:
\begin{equation}\label{a6}
a =
a_0\left( 1+\cot(\frac{\pi}{8}+\Omega+\delta)\right)%
\end{equation}
where $a_0$ is

\begin{equation} \label{a0vdW}
a_0=R_{vdW}\frac{\Gamma (3/4)}{2\sqrt{2}\Gamma (5/4)}\simeq
4.99\mbox{ a.u.}
\end{equation}
and $\Omega=19.38$ is the WKB phase (\ref{Omega}), accumulated in the domain $R_r<R<\infty$.

To  calculate the   Jost function, we determine   the variation of the scattering
length (\ref{a6}) with the small variation of phase $\Delta\delta$:
\begin{equation}\label{Delta6}
\Delta a=-a_0\frac{\Delta \delta}{\sin^2\left(\frac{\pi}{8}+\Omega+\delta\right)}.
\end{equation}
In the case of the zero-range potential
\begin{equation}\label{ZRP0}
V_{ZRP}=\frac{2\pi}{M} a_{sc} \delta(\textbf{R})(\frac{\partial}{\partial R}R)
\end{equation}
the variation of phase is given by (\ref{deltStCoul}): 
\begin{equation}
\Delta \delta=-2\pi M a_{sc}.
\end{equation}

The corresponding variation of the scattering amplitude, induced  by the zero-range
potential (\ref{ZRP0}),
can be obtained within the distorted wave approximation:
\begin{equation}\label{DWA}
\Delta f=-\frac{M}{2\pi} \int \Psi_0(\textbf{R})^*V_{ZRP}\Psi_0(\textbf{R}) d^3R=-a_{sc}|\Psi_0(0)|^2
\end{equation}
where $\Psi_0(\textbf{R})$ is the wave function calculated  without the account
of the  potential (\ref{ZRP0}).
We now use eq. (\ref{Psi0}) to express the wave function in terms of the Jost function and obtain
\begin{equation}
\label{Deltaf}
\Delta f=-\frac{a_{sc}}{|f_J(0)|^2}.
\end{equation}
The $S$-wave scattering length is connected to the zero-energy scattering amplitude by the
relation  $a=-f(0)$.
Inserting this relation into eq. (\ref{Deltaf}) and comparing (\ref{Deltaf}) with  (\ref{Delta6}),  we finally obtain  the following expression for $1/|f_J(0)|^2$:
\begin{equation}
\label{JF}
1/|f_J(0)|^2=\frac{2\pi M a_0}{\sin^2(\pi/8+\Omega+\delta)}.
\end{equation}
\vfill \eject

\bibliographystyle{unsrt}
\bibliography{hahspinex}




\end{document}